\documentclass[sigconf]{acmart}

\usepackage{booktabs} 
\usepackage{todonotes}
\usepackage{microtype}
\usepackage{xspace}
\usepackage{balance}
\usepackage{algorithm}
\usepackage{algpseudocode}
\usepackage{caption}
\usepackage{multicol} 
\usepackage{subfig}
\usepackage[capitalise]{cleveref}

\usepackage{pifont}

\usepackage{tikz}
\usepackage{pgflibraryshapes}
\usetikzlibrary{arrows}
\usetikzlibrary{chains}
\usetikzlibrary{fadings}
\usetikzlibrary{matrix}
\usetikzlibrary{fit}
\usetikzlibrary{positioning}
\usetikzlibrary{shapes}
\usetikzlibrary{automata}
\tikzset{
 ctrlstate/.style = {state,align=center,inner sep=2pt, minimum size=2mm},
 cfastate/.style = {ctrlstate},
 cfatargetstate/.style = {cfastate, double},
 compstate/.style = {cfastate, rounded rectangle, minimum height=5mm, minimum width=3em, inner sep=3pt},
 concept/.style = {cfastate, inner sep=3pt, fill=gray!50, rectangle,
        minimum width=17mm, minimum height=9mm, draw=gray!6 },
 crosscutting/.style = {concept, rounded rectangle, fill=gray!30},
 conceptstate/.style = {concept, rounded rectangle, fill=gray!30, draw=gray!90, minimum width=20mm},
 inputstate/.style = {concept, rectangle, fill=gray!10, draw=gray!90, minimum width=20mm, inner sep=3pt},
 explain/.style = {circle, draw=gray!30, line width=1mm, minimum size=7mm},
 one/.style = {fill=blue!70,draw=blue!70},
 two/.style = {fill=red!50,draw=red!50},
 abststate/.style = {rectangle,align=center,inner sep=2pt,minimum size=3.5mm,fill=gray!0, draw=gray!90},
 line/.style = {draw},
 trans/.style = {draw,semithick,->,shorten >=1pt,>=stealth'},
 missing/.style = {draw=red,densely dotted,fill=red,semithick,->,shorten >=1pt,>=stealth'},
 ctrans/.style = {draw,very thick,->,shorten >=1pt,>=stealth',draw=gray!90},
 epsilon/.style = {trans,dashed},
 strengthen/.style = {draw=gray!30,semithick,double,shorten >=1pt,>=stealth',line width=1mm},
}

\newcommand{\motionblockleft}{\begin{mbox}\sf\begin{tikz}[baseline=(X.base)]\node[draw=black!60,fill=blue!12,semithick,rectangle,inner sep=1pt, minimum size=1em, outer sep=0pt, rounded corners=1pt] (X)}%
\newcommand{\controlblockleft}{\begin{mbox}\sf\begin{tikz}[baseline=(X.base)]\node[draw=black!60,fill=orange!15,semithick,rectangle,inner sep=1pt, minimum size=1em, outer sep=0pt, rounded corners=1pt] (X)}%
\newcommand{\hatblockleft}{\begin{mbox}\sf\begin{tikz}[baseline=(X.base)]\node[draw=black!60,fill=yellow!20,semithick,rectangle,inner sep=1pt, minimum size=1em, outer sep=0pt, rounded corners=1pt] (X)}%
\newcommand{\looksblockleft}{\begin{mbox}\sf\begin{tikz}[baseline=(X.base)]\node[draw=black!60,fill=violet!20,semithick,rectangle,inner sep=1pt, minimum size=1em, outer sep=0pt, rounded corners=1pt] (X)}%
\newcommand{\sensingblockleft}{\begin{mbox}\sf\begin{tikz}[baseline=(X.base)]\node[draw=black!60,fill=cyan!20,semithick,rectangle,inner sep=1pt, minimum size=1em, outer sep=0pt, rounded corners=1pt] (X)}%
\newcommand{\soundblockleft}{\begin{mbox}\sf\begin{tikz}[baseline=(X.base)]\node[draw=black!60,fill=magenta!20,semithick,rectangle,inner sep=1pt, minimum size=1em, outer sep=0pt, rounded corners=1pt] (X)}%
\newcommand{\operatorblockleft}{\begin{mbox}\sf\begin{tikz}[baseline=(X.base)]\node[draw=black!60,fill=green!20,semithick,rectangle,inner sep=1pt, minimum size=1em, outer sep=0pt, rounded corners=1pt] (X)}%

\newcommand{\blockleft}{\begin{mbox}\sf\begin{tikz}[baseline=(X.base)]\node[draw=black!60,fill=black!3,semithick,rectangle,inner sep=1pt, minimum size=1em, outer sep=0pt, rounded corners=1pt] (X)}%
\newcommand{\blockright}{;\end{tikz}\normalfont\end{mbox}}%

\newcommand{\mblockleft}{\begin{mbox}\sf\begin{tikz}[baseline=(X.base)]\node[draw=red!60,densely dotted,fill=red!3,semithick,rectangle,inner sep=1pt, minimum size=1em, outer sep=0pt, rounded corners=1pt] (X)}%
\newcommand{\mblockright}{;\end{tikz}\normalfont\end{mbox}}%

\renewcommand{\paragraph}[1]{\vspace{0.08em}\noindent {\bf #1}}

\AtBeginDocument{%
  \providecommand\BibTeX{{%
    \normalfont B\kern-0.5em{\scshape i\kern-0.25em b}\kern-0.8em\TeX}}}

\copyrightyear{2022}
\acmYear{2022}
\setcopyright{acmcopyright}\acmConference[WiPSCE '22]{Proceedings of the 17th Workshop in Primary and Secondary Computing Education}{October 31-November 2, 2022}{Morschach, Switzerland}
\acmBooktitle{Proceedings of the 17th Workshop in Primary and Secondary Computing Education (WiPSCE '22), October 31-November 2, 2022, Morschach, Switzerland}
\acmPrice{15.00}
\acmDOI{10.1145/3556787.3556859}
\acmISBN{978-1-4503-9853-4/22/10}

\newcommand{\summary}[2]{
	\vspace{0.4em}
	\noindent
	\colorbox{gray!20}{%
		\parbox{.97\linewidth}{%
			\textbf{\textsf{Summary (\textit{#1})}}
			#2
		}%
	}%
}%

\newcommand{\numprojectsBugpattern}{1,903\xspace}
\newcommand{\numBugpattern}{6,129\xspace}
\newcommand{\numprojectsSmell}{157\xspace}
\newcommand{\numSmell}{592\xspace}
\newcommand{\numprojectsPerfumes}{2,284\xspace}
\newcommand{\numPerfumes}{14,495\xspace}

\newcommand{\classicbugpatternmblock}{4,942\xspace}
\newcommand{\classicbugpatternscratch}{18,698\xspace}
\newcommand{\pvalueclassicbugpattern}{<0.001\xspace}
\newcommand{\aclassicbugpattern}{0.33\xspace}
\newcommand{\allbugpatternmblock}{11,071\xspace}
\newcommand{\allbugpatternscratch}{18,698\xspace}
\newcommand{\pvalueallbugpattern}{0.126\xspace}

\newcommand{\classicsmellmblock}{23,933\xspace}
\newcommand{\classicsmellscratch}{43,996\xspace}
\newcommand{\pvalueclassicsmell}{<0.001\xspace}
\newcommand{\aclassicsmell}{0.43\xspace}
\newcommand{\allsmellmblock}{24,525\xspace}
\newcommand{\allsmellscratch}{43,996\xspace}
\newcommand{\pvalueallsmell}{<0.001\xspace}
\newcommand{\aallsmell}{0.43\xspace}
\newcommand{\classicperfumemblock}{30,296\xspace}
\newcommand{\classicperfumescratch}{135,494\xspace}
\newcommand{\pvalueclassicperfume}{<0.001\xspace}
\newcommand{\aclassicperfume}{0.34\xspace}
\newcommand{\allperfumemblock}{44,791\xspace}
\newcommand{\allperfumescratch}{135,494\xspace}
\newcommand{\pvalueallperfume}{<0.001\xspace}
\newcommand{\aallperfume}{0.45\xspace}

\newcommand{\pvaluemostcomplexscript}{0.4\xspace}

\newcommand{\pvalueclassicbupgatternperblockcount}{<0.001\xspace}
\newcommand{\aclassicbupgatternperblockcount}{0.32\xspace}

\newcommand{\pvalueclassicsmellperblockcount}{0.003\xspace}
\newcommand{\aclassicsmellperblockcount}{0.48\xspace}

\newcommand{\pvalueclassicperfumeperblockcount}{<0.001\xspace}
\newcommand{\aclassicperfumeperblockcount}{0.29\xspace}

\newcommand{\maxblocksmblock}{2,923\xspace}
\newcommand{\maxblocksscratch}{6,790\xspace}
\newcommand{\numMcoreProj}{3,023\xspace}
\newcommand{\numCodeyProj}{529\xspace}
\newcommand{\countbothrobots}{27\xspace}

\begin{document}

\title{Common Patterns in Block-Based Robot Programs}

\author{\mbox{Florian Obermüller}}
\affiliation{%
	\institution{University of Passau}
	\city{Passau}
	\country{Germany}
}

\author{Robert Pernerstorfer}
\affiliation{%
	\institution{University of Passau}
	\city{Passau}%
	\country{Germany}
}

\author{Lisa Bailey}
\affiliation{%
	\institution{University of Passau}
	\city{Passau}%
	\country{Germany}
}

\author{Ute Heuer}
\affiliation{%
	\institution{University of Passau}
	\city{Passau}%
	\country{Germany}
}

\author{Gordon Fraser}
\affiliation{%
	\institution{University of Passau}
	\city{Passau}
	\country{Germany}
}
\renewcommand{\shortauthors}{Obermüller, et al.}

\begin{abstract}
	Programmable robots are engaging and fun to play with, interact with the
real world, and are therefore well suited to introduce young learners to
programming.
	Introductory robot programming languages often extend existing block-based
languages such as \Scratch. While teaching programming with such languages is
well established, the interaction with the real world in robot programs leads
to specific challenges, for which learners and educators may require assistance
and feedback.
	A practical approach to provide this feedback is by identifying and
pointing out patterns in the code that are indicative of good or bad solutions.
While such patterns have been defined for regular block-based programs,
robot-specific programming aspects have not been considered so far. The aim of
this paper is therefore to identify patterns specific to robot programming for
the \Scratch-based \mblock programming language, which is used for the popular \mbot and
\codeyrocky robots.
	We identify: (1)~\numbugpatterns~\emph{bug patterns}, which indicate erroneous code; (2)
three \emph{code smells}, which indicate code that may work but is written in a
confusing or difficult to understand way; and (3) \numperfumes \emph{code perfumes},
which indicate aspects of code that are likely good.
	We extend the \litterbox analysis framework to automatically identify these
patterns in \mblock programs. Evaluated on a dataset of \numRobotProj \mblock
programs, we find a total of \numBugpattern instances of bug patterns,
\numSmell code smells and \numPerfumes code perfumes.
	This demonstrates the potential of our approach to provide feedback and
assistance to learners and educators alike for their \mblock robot programs.
\end{abstract}



\newcommand{\litterbox}{\textsc{LitterBox}\xspace}
\newcommand{\Scratch}{\textsc{Scratch}\xspace}
\newcommand{\drscratch}{\textsc{Dr. Scratch}\xspace}
\newcommand{\hairball}{\textsc{Hairball}\xspace}
\newcommand{\qualityhound}{\textsc{Quality Hound}\xspace}
\newcommand{\findbugs}{\textsc{FindBugs}\xspace}
\newcommand{\catnip}{\textsc{Catnip}\xspace}
\newcommand{\whisker}{\textsc{Whisker}\xspace}
\newcommand{\bastet}{\textsc{Bastet}\xspace}
\newcommand{\itch}{\textsc{Itch}\xspace}
\newcommand{\mbot}{\textsc{mBot}\xspace}
\newcommand{\codeyrocky}{\textsc{Codey Rocky}\xspace}
\newcommand{\mindstormsev}{\textsc{Mindstorms EV3}\xspace}
\newcommand{\wedo}{\textsc{WeDo}\xspace}
\newcommand{\boost}{\textsc{Boost}\xspace}
\newcommand{\Ozobot}{\textsc{Ozobot}\xspace}

\newcommand{\mblock}{\textsc{mBlock}\xspace}
\newcommand{\makeblock}{\textsc{MakeBlock}\xspace}
\newcommand{\rockylight}{\textsc{Rocky Light}\xspace}

\newcommand{\codey}{\texttt{codey}\xspace}
\newcommand{\mcore}{\texttt{mcore}\xspace}
\newcommand{\json}{\texttt{JSON}\xspace}
\newcommand{\pjson}{\texttt{.JSON}\xspace}
\newcommand{\issue}{\texttt{Issue}\xspace}
\newcommand{\issues}{\texttt{Issues}\xspace}
\newcommand{\opcode}{\texttt{Opcode}\xspace}
\newcommand{\opcodes}{\texttt{Opcodes}\xspace}

\newcommand{\numMBlock}{28,192\xspace}
\newcommand{\numRobot}{4,126\xspace}
\newcommand{\numRobotProj}{3,540\xspace}
\newcommand{\numCodey}{1,020\xspace}
\newcommand{\numMcore}{3,106\xspace}
\newcommand{\numMBlockBugs}{4,541\xspace}
\newcommand{\numMBlockBugProj}{1,740\xspace}
\newcommand{\numAllBugs}{122,844\xspace}
\newcommand{\numAllBugProj}{3,366\xspace}

\newcommand{\numclassified}{164\xspace}
\newcommand{\numfailure}{137\xspace}
\newcommand{\numfalsepositive}{12\xspace}
\newcommand{\numunreachable}{5\xspace}
\newcommand{\numnotnoticable}{10\xspace}
\newcommand{\numbugpatterns}{26\xspace}
\newcommand{\numbugpatternsfound}{19\xspace}
\newcommand{\numsmells}{3\xspace}
\newcommand{\numperfumes}{18\xspace}
\newcommand{\numperfumesfound}{16\xspace}

\maketitle

\section{Introduction}\label{sec:intro}

\begin{figure}[t]
	\centering
	\includegraphics[width=0.7\columnwidth]{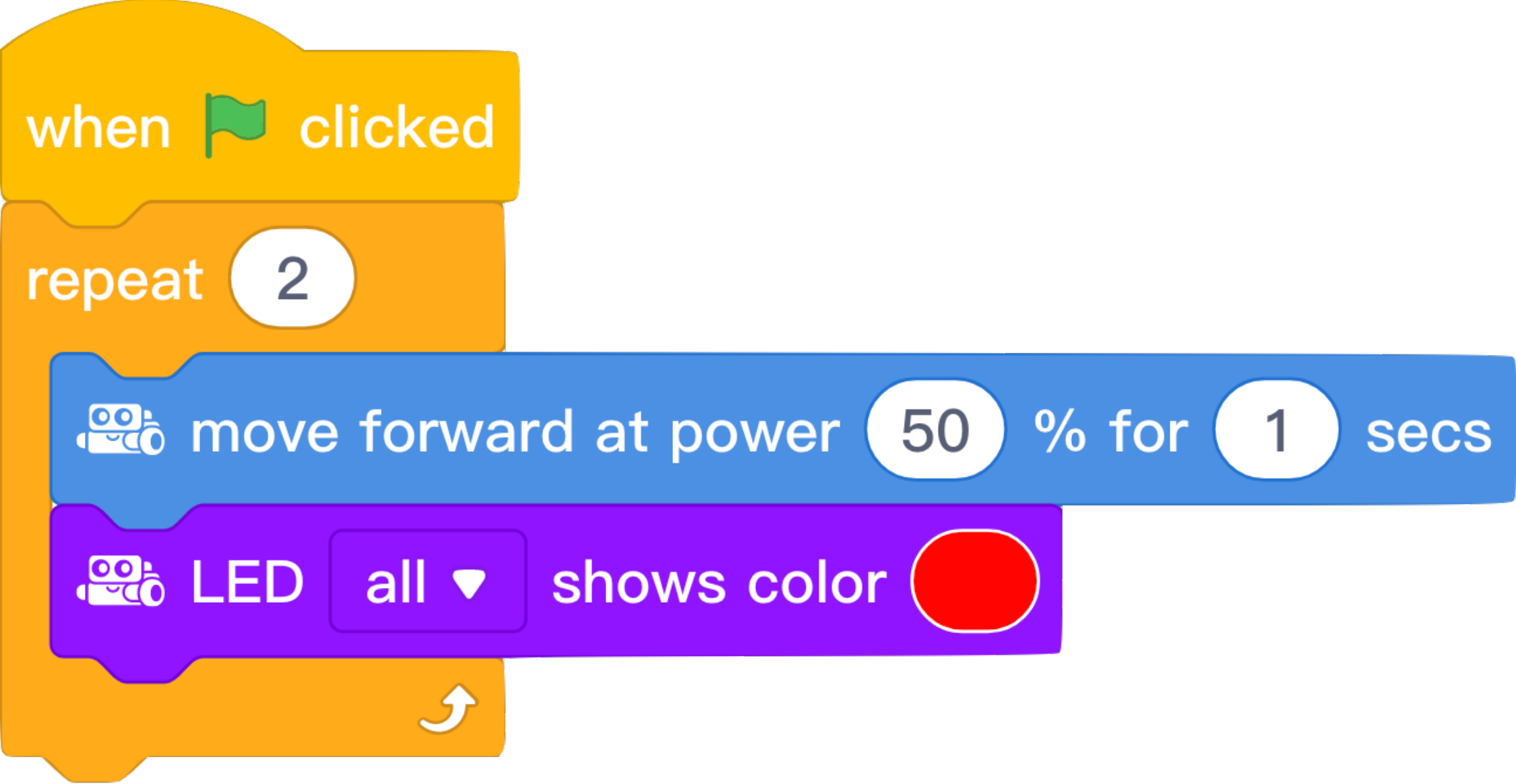} \\
\vspace{1em}
\noindent\fcolorbox{black}{black!5}{\parbox{.95\columnwidth}{
    \textbf{Bug Pattern LED Off Missing found}: The LEDs on your robot are still turned on after the program has stopped. Add an LED Off block at the end of your program.
    }}    
	\caption{\label{fig:example_bug}Example \mblock program for an \mbot robot, containing a bug: The robot is moved forward twice and the LEDs are set to red, but they are not turned off again. This bug is identified by our \litterbox extension and provides a textual hint.}
        
\end{figure}

A common means to introduce young learners to programming is by using robots.
Programmable robots are fun to interact with and engage learners. Their interactions with the real world through sensors and actuators rather than simulated environments make them well suited for cross-curricular activities.
The programming environments commonly used for controlling such robots are based on
established introductory block-based programming languages. For example, the
popular \Scratch\cite{maloney2010} programming environment offers `extensions' that add dedicated
blocks for controlling different types of Lego robots (e.g., \mindstormsev,
\boost, \wedo), and the various robots produced by \makeblock (e.g., \mbot, \codeyrocky) can be programmed in \mblock, a modified version of \Scratch. These
modified programming environments thus offer intuitive first steps into entry
level programming and make the transition to purely computer-based programming
easy~\cite{bau2017}.

\Cref{fig:example_bug} shows an example \mblock program controlling an \mbot
robot: When the program is started on the computer, the robot is moved forward twice for a second at half its maximum speed, and its LEDs are set to
red. Except for the two robot-specific blocks, the program itself is
indistinguishable from any other standard \Scratch program learners might
create. However, the robot-specific blocks have real-world implications that
learners need to know and comprehend in order to properly control the robots. For
example, while the `move forward' block turns off the motor after one second,
the `LED' block only turns the lights on, but not off. Consequently, the program has
a bug: Executing the code will have the side-effect that the lights are
constantly on while the program is running, and also remain turned on even after the execution has completed.
This bug is based on a common misunderstanding of learners who start to interact with
robots.

To support learners in overcoming their misconceptions and building correctly
working programs, they may require assistance. While this assistance by default
is provided by educators, these might be overwhelmed in a classroom of
students, each with individual problems in their programs. However, many
challenges in programming are repetitive and can be detected automatically in
learners' programs. Different types of patterns can be identified: \emph{Bug
patterns} are code constructs that provide evidence of misconceptions and bugs;
\emph{code smells} are aspects of code that may achieve the desired output, but
are ineffective or confusing in some way; and \emph{code perfumes} provide
evidence of correctly applied code constructs and idioms. For block-based
programming languages like \Scratch, tools identifying these types of patterns
are available~\cite{fraser2021, obermueller2021perfumes}. 
However, existing tools and patterns do not cover the specific
challenges caused by robot programming.

\looseness=-1
In this paper, we therefore aim to extend the concept of these patterns to
introductory robot programming. We create a dataset of \numRobotProj programs
written in the \mblock programming language for the popular \mbot and
\codeyrocky programmable robots. Using this dataset, we determine and evaluate a
catalogue of bug patterns, code smells, and code perfumes, and implement them
as an \mblock extension of the \litterbox\cite{fraser2021} analysis framework.
When instances of the negative code patterns are found, our implementation also
automatically generates hints, providing feedback on why there is a problem and what the underlying misconception may be. Furthermore our automatically 
generated hints also suggest how to fix the bug or to remove the code smell.
For the program in \Cref{fig:example_bug}, our \litterbox extension reports an
instance of the \emph{LED Off Missing} bug pattern, stating that the LEDs are turned on 
after the program ended and that the user should introduce an LED Off block to fix 
this problem.

Our experiments suggest that \Scratch projects are structurally different from
\mblock programs; they are smaller and have fewer scripts, but individual
scripts can be similarly complex. We found instances for most bug patterns,
code smells and code perfumes in our dataset, suggesting that the patterns are
highly applicable. A manual classification also demonstrates that \mblock bug
patterns are a frequent cause of failures, i.e., program states where the robot observably misbehaves. The integration of these patterns
into \litterbox enables educators and learners to immediately make use of this
information.

\section{Background}\label{sec:background}

In order to engage young learners with the concepts of programming, two
important approaches are (1) to simplify the construction of programs using
block-based languages and (2) to use programmable robots. Constructing such programs for robots gives rise to
specific challenges. The aim of
this paper is to derive and evaluate code patterns that can help to address
these challenges.

\subsection{Educational Robots}\label{subsec:robots}

\looseness=-1 
A popular way to introduce children to programming is using programmable
robots~\cite{mcgill2020}. This has multiple reasons:
First, robots offer an easy starting point as they usually can be controlled without a computer, such as the \Ozobot robots~\cite{koerber2021}. 
Second, interacting with the environment rather than programming simulated environments on the computer can lead to higher learning motivation~\cite{peng2020study}: Students have to investigate aspects of the robots' capabilities and can then tackle real world problems, like reading from a sensor and letting the robot act accordingly.
Third, programming robots leads to a combination of acquiring programming skills with other abilities like spatial thinking~\cite{jung2018}, and the use of robots may also lead to further discussion about the consequences of programming and program execution, for example as motors could be overstrained or other parts of the robot could be damaged.
Finally, robots are well suited for cross-curricular activities (e.g., physics,
art, physical education) due to their sensors and
actuators~\cite{sullivan2016robotics}. As an example, the workings
of an ultra sonic sensor can be discussed, and measured data can be used in classic
tasks like calculating the speed of the robot from time and travelled way.

Educational robot programming environments are usually intended to help 
transition to solely computer based programming. For example, the
\makeblock line of robots and their
\mblock\footnote{\url{http://mblock.makeblock.com/}, last accessed 01.06.22}
programming environment achieve this by using an extended version of
the popular block based programming language \Scratch~\cite{maloney2010}.
\mblock uses the exact same blocks and shapes as \Scratch to prevent syntactical
errors for making programming more accessible for novices. The
\Scratch programming environment is extended with new blocks for controlling the robots'
actuators and reading the sensors of the robots when connected to the computer. 

\looseness=-1 
Two popular types of robots compatible with \mblock are the \codeyrocky and the
\mbot. Both robots have two motors to move each side separately. Also, with both robots
one can sense the intensity of the ambient light, display information on an LED matrix, and turn
LEDs off and on. Furthermore, the \mbot has an ultra sonic sensor for measuring
distances and a line following sensor to detect if the robot is driving over
dark or bright ground. The \codeyrocky, on the other hand, has a gyroscope,
additional lights, a colour sensor as well as a potentiometer. Thus, both
robots have a variety of sensors and actuators that can be read and controlled with the
additional blocks of \mblock. 

\subsection{Patterns in Block-based Programs}\label{subsec:block_based_patterns}
\looseness=-1
Even though block-based languages are designed with the aim to make programming
more accessible and intuitive to novices, it can nevertheless be challenging to
assemble the blocks in a correct way that implements the desired functionality.
An important means to provide feedback and support to learners, and for
research or educational purposes, is to identify common patterns of blocks in
the learners' programs. There are three main categories of such patterns which
have been explored in the context of block-based programs.

\looseness=-1 
\emph{Code smells} are idioms of code that decrease the understandability of
the project and increase the likelihood of bugs occurring when modifying the
code~\cite{fowler1999}. A range of studies have investigated the types and
occurrences of code smells in block-based
languages~\cite{hermans2016a,maloney2010,moreno2014,techapalokul2017a}. There
is evidence that the presence of code smells hampers the ability of learners to
modify the code~\cite{hermans2016b}, and that code smells can decrease the
likelihood of projects being reused~\cite{techapalokul2017a} in
remixes~\cite{bau2017}. Code smells can be detected automatically using tools
such as \qualityhound~\cite{techapaloku2017b}, \hairball~\cite{boe2013}, or
\litterbox~\cite{fraser2021}.

\looseness=-1 
While code smells capture attributes that are independent of whether the
affected code is correct or not, \emph{bug patterns} refer to aspects of code
that are likely to lead to undesired behaviour (i.e., \emph{bugs} or
\emph{defects}~\cite{hovemeyer2004}). Similar to code smells, bug patterns can
be detected automatically on source
code~\cite{lintersGeneral,codeAnalysisTools}. Bug patterns have been shown to
appear frequently in \Scratch projects~\cite{fraedrich2020}, and there are
automated tools that can be used to find them, such as
\litterbox~\cite{fraser2021}.
Note that instances of bug patterns in code are likely candidates of bugs, but are not guaranteed to be incorrect. Determining whether a program is truly broken would require running the program and testing whether it behaves as expected. While there are approaches to do this automatically also for block-based languages (e.g., \whisker~\cite{stahlbauer2019testing}, \itch~\cite{johnson2016itch}, \bastet~\cite{stahlbauer2020}), these approaches require task-specific tests or specifications that describe the expected behaviour. In contrast, patterns are task-independent, and therefore only need to be defined once in order to enable detection on any program. 

\looseness=-1 
Code smells and bug patterns aim to find problems, but it is
also possible to identify positive aspects of code, which may for example serve
as evidence of progress or as positive feedback. Code idioms
indicating good programming practices or code showing understanding of certain
programming concepts are known as \emph{code
perfumes}~\cite{obermueller2021perfumes}, and can be seen as the opposite of
code smells. Technically, matching code perfumes on block-based programs is
similar to matching code smells and bug patterns, which means tools like
\litterbox~\cite{fraser2021} can also detect code perfumes automatically.

While these concepts of code patterns are well explored in the context of
block-based programming, none of the existing approaches focus explicitly on
the robot-specific aspects of code. The aim of this paper is to fill this gap.
Since \mblock is based on the \Scratch programming language and extends it by the
robot functionality, we expect to find new types of all statically detectable
code patterns in the programs for \mbot and \codeyrocky robots.

\section{Patterns in \mblock}
\label{sec:patterns}

Along with the additional possibility of controlling robots, and thus real physical hardware, new problems can occur. 
These problems can arise from the physical limitations of the robots, from the way the robot software itself is implemented, or also from peculiarities in handling robots or physical hardware, e.g., reading out sensor values and reacting to them. 
Furthermore, there are also new situations where learners can profit from positive feedback.
By reflections of students of computer science education, programming courses with children, comparing robot behaviour with known patterns in \Scratch, and by our own experimentation with the robots, we discovered 26 bug patterns, three code smells, and 18 code perfumes.

\subsection{Bug Patterns}
A bug pattern in \Scratch is a composition of blocks that are typical of buggy code, or a general buggy deviation from a correct code idiom~\cite{fraedrich2020}.
Following this notion, robot bug patterns in \mblock are compositions of blocks that cause the robot to act incorrectly or exhibit undesirable behaviour.

\begin{figure}[t]
	\centering
	\includegraphics[scale=0.3]{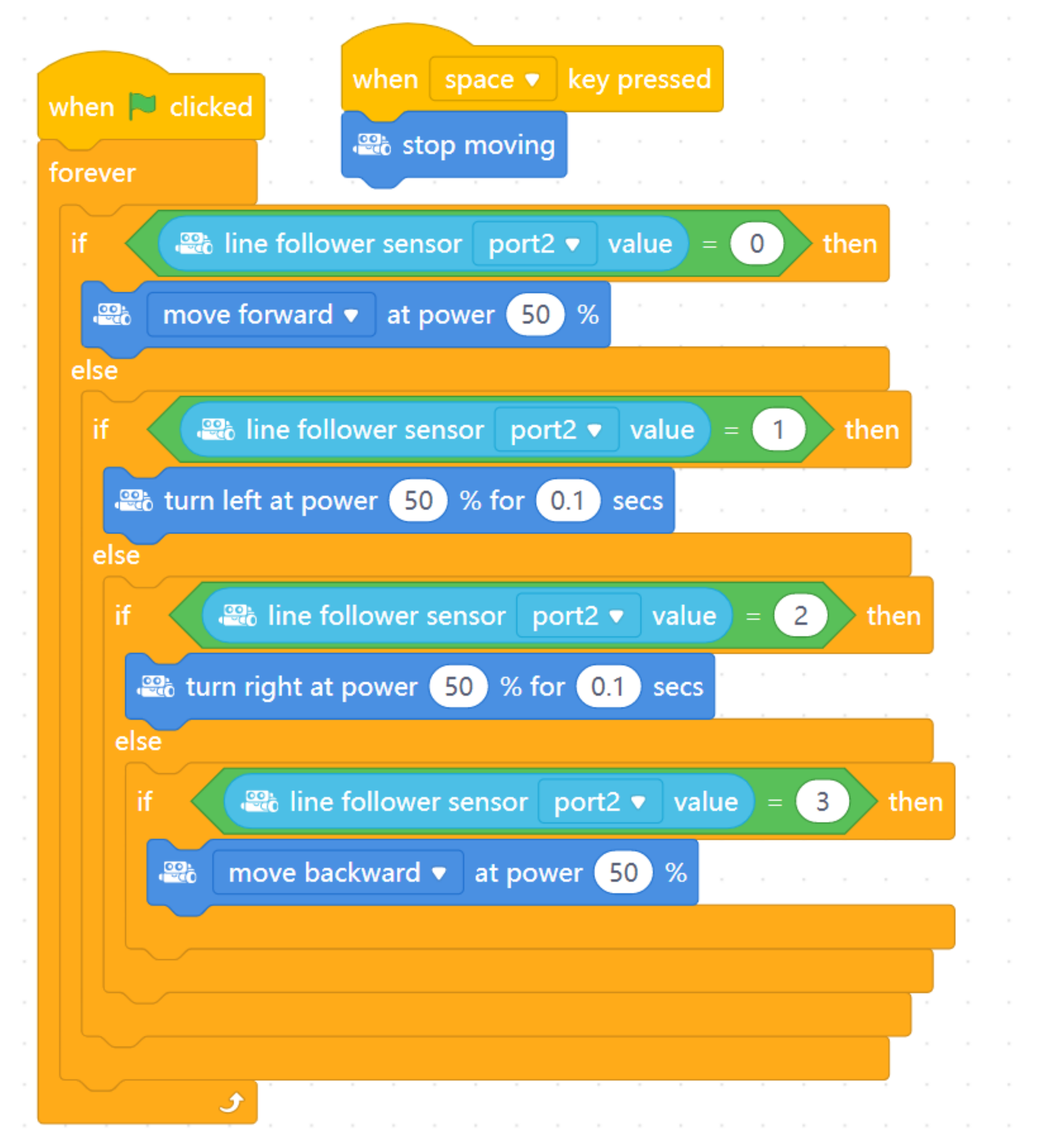}
	\vspace{-1em}
	\caption{\label{fig:bugpattern_04}Example Action Not Stopped bug pattern.}
\end{figure}

\paragraph{Action Not Stopped:}
Stop commands for actuators (LED, light, matrix, motor) only switch off the corresponding actuator, but do not end the scripts in which the actuator is used. The scripts themself, possibly containing loops, continue to run. 
If a statement for an actuator is within a loop and this actuator is terminated by a stop command from another script, the statement will nevertheless be executed again by the loop.
If all uses of an actuator are to be stopped, the scripts that use this actuator have to be stopped, too. \Cref{fig:bugpattern_04} shows an example of this bug pattern, where the actuators are not correctly stopped.

\paragraph{Actuator Deactivation Missing:}
In contrast to time-limited statements, some blocks only switch actuators on without switching them back off after a while. Even after all scripts have stopped, the actuators remain active. 
If the program code contains no scripts that turn them off permanently, the only possibility to deactivate them is by completely switching off the robot. 
This problem can easily be prevented with scripts that only switch on the actuators for a limited time, or by a separate stop script that deactivates them. We define several versions of this bug pattern depending on the specific actuator used:
\begin{itemize}
\item \textbf{LED Off Missing}	
\item \textbf{Light Off Missing}
\item \textbf{Matrix Off Missing}
\item \textbf{Motor Off Missing}
\end{itemize}
\Cref{fig:example_bug} is an example of the LED Off Missing bug pattern.

\paragraph{Colour Out of Range:}
The defined values for colours are the integers from 0 to 255. Setting the colour to other values is possible, but values higher than 255 lead to an actual colour setting of 255, and negative values to 0. Thus, the code would not match the result, and only integers form 0 to 255 should be used.

\paragraph{Interrupted Loop Sensing:}
A typical concept of robot programming is making the robot react to specific sensor values. Using a sensor query within a loop that contains time-limited statements can cause a bug: When the sensor values are not read out frequently enough, the robot might not react to short occurrences of the relevant values. To prevent this, time-limited statements and queries concerning sensors should be in separate loops.

\paragraph{Low Motor Power:} 
The electric motors of the \mbot robot need a minimum amount of power to move
the wheels. When the power value of a movement statement
is set to less than 25\%, the corresponding wheels will not turn. Programs targeting
low movement speeds should be avoided or used for a \codeyrocky robot instead, which does not have a minimum threshold for the motor power.

\paragraph{Missing Loop Sensing:}
In order to make the robot react to a specific change in the values of a sensor, one must continuously read out the corresponding sensor values. When a query concerning sensors is not within a loop, it will only be executed once, which leads to the robot not reacting to later value changes. In order to read out the sensor values continuously, the query should be inside a forever loop or a loop with a stopping condition.

\begin{figure}[t]
	\centering
	\includegraphics[scale=0.25]{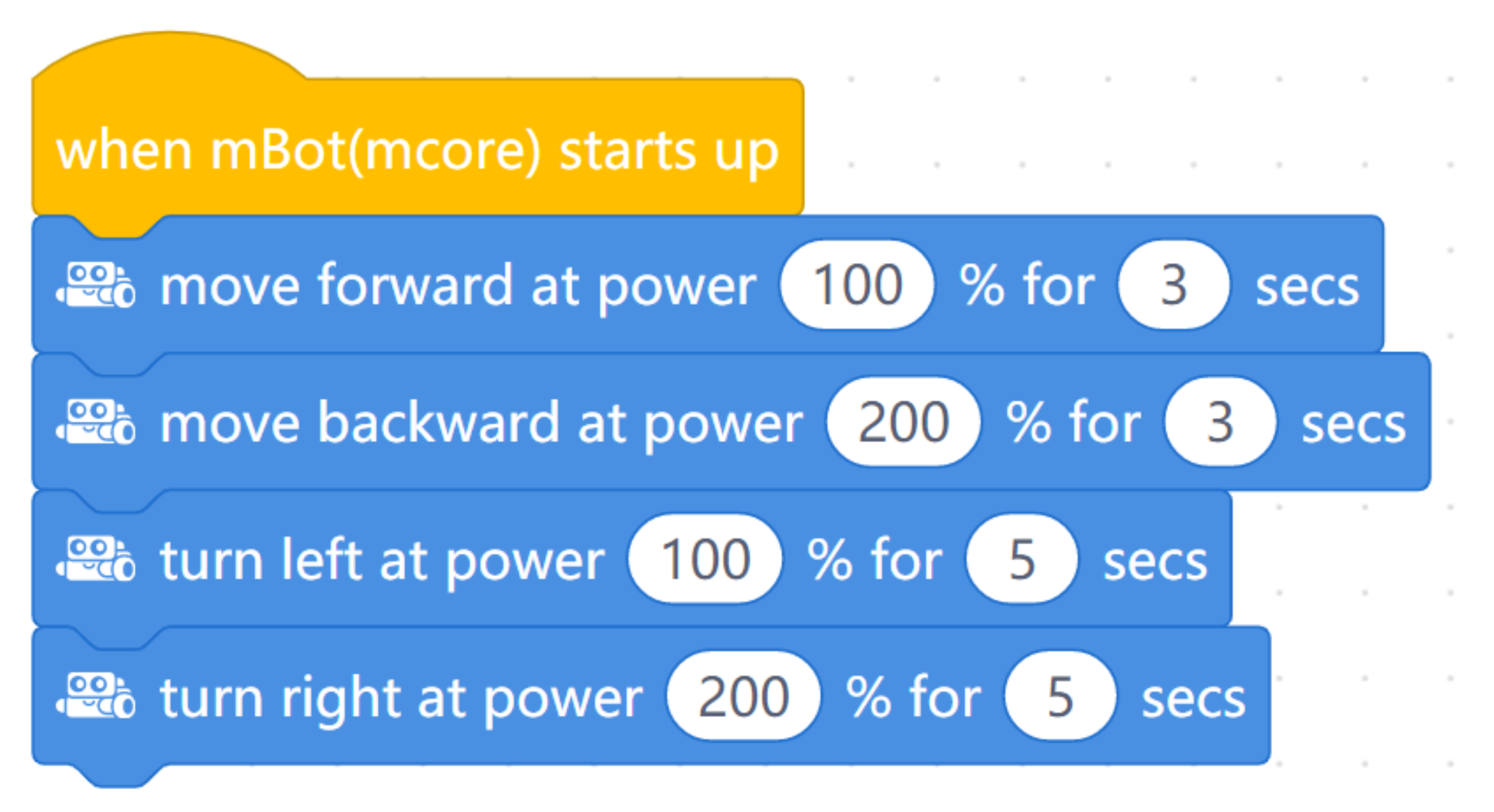}
	\vspace{-1em}
	\caption{\label{fig:codesmell_13}Example Motor Setting Out Of Range bug pattern.}
	\vspace{-1em}
\end{figure}

\paragraph{Motor Out Of Range:}
The motors of the robots can only be controlled with a maximum of
100\%. It is possible to set higher power values, but in practice, the
motor still does not run faster.  When power values above 100\% are
used, the result does not match the code, since all values above 100\%
lead to the same speed as 100\%. Analogously, this also applies to
values beneath -100\%.  In general, only power values within a
reasonable range (i.e., up to 100\%) should be
used. \Cref{fig:codesmell_13} shows an example of this bug pattern.

\begin{figure}[t]
	\centering
	\includegraphics[scale=0.25]{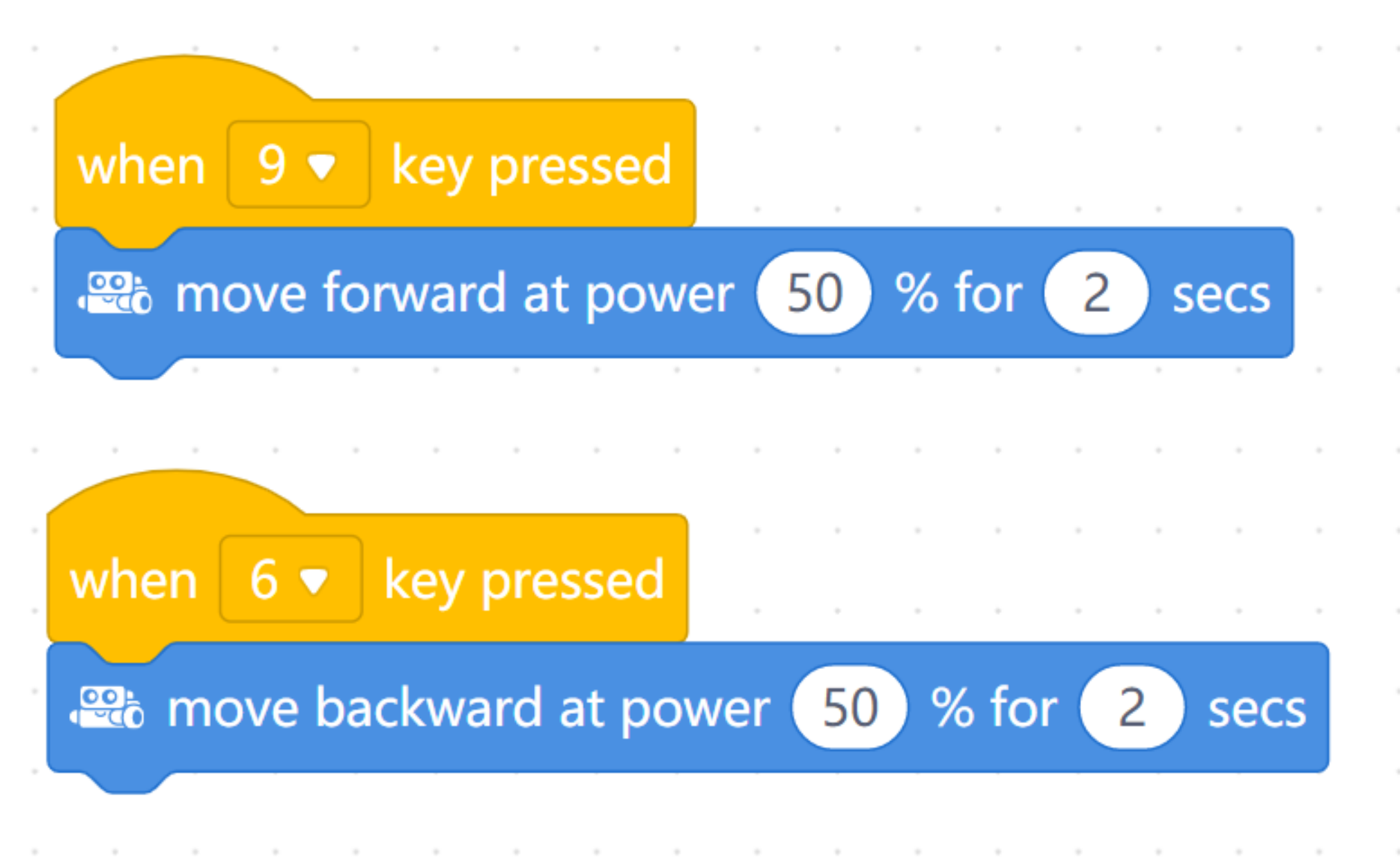}
	\vspace{-1em}
	\caption{\label{fig:bugpattern_09}Example Parallel Actuator Use bug pattern.}
\end{figure}

\paragraph{Parallel Actuator Use:} 
When two scripts run in parallel and use the same actuators, they block each
other or cancel the other use. To avoid such conflicts, different
scripts should use different actuators, or one has to make sure that other scripts have
finished beforehand. An example of this bug pattern is shown in
\Cref{fig:bugpattern_09}.

\begin{figure}[t]
	\centering
	\includegraphics[scale=0.3]{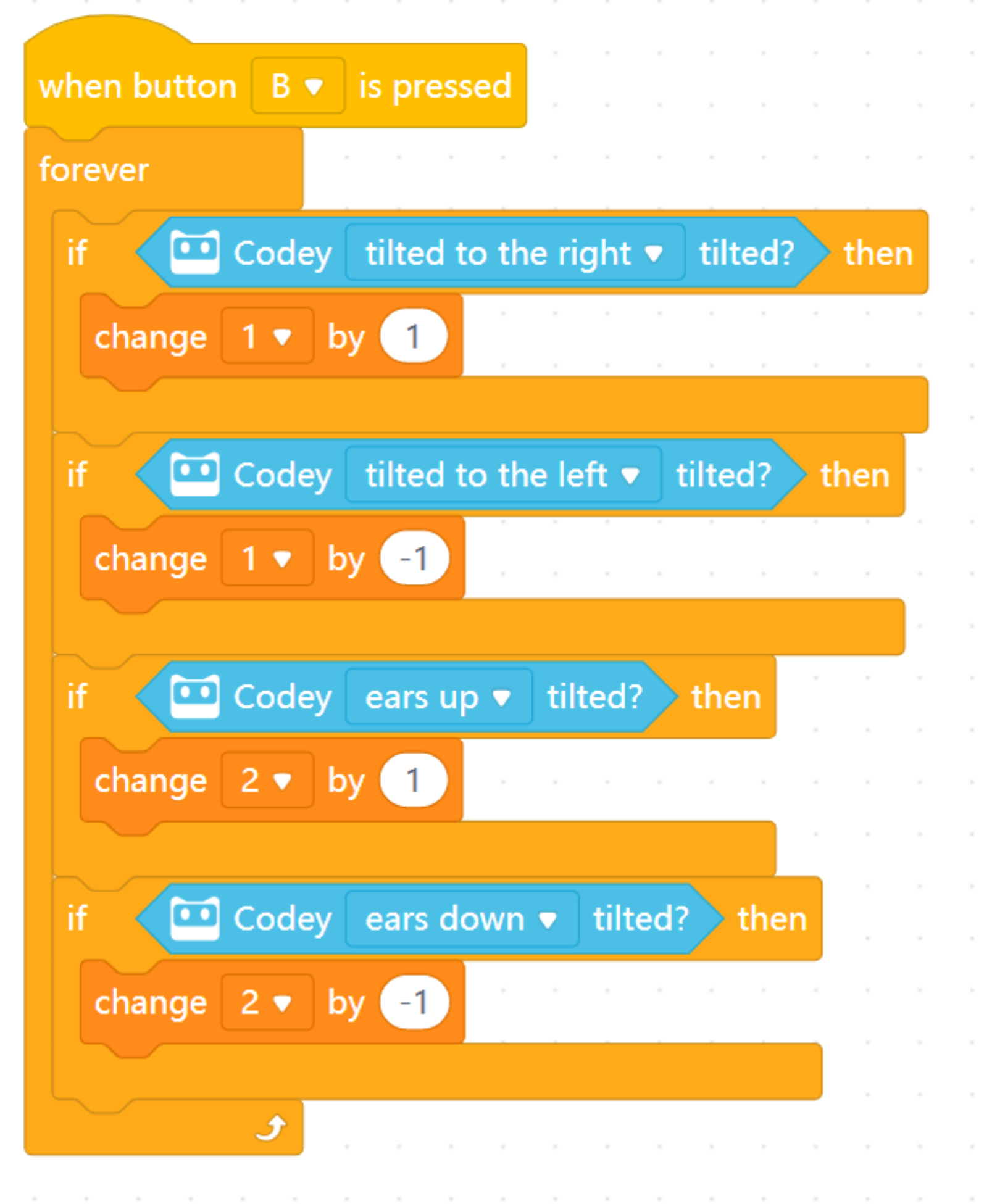}
	\vspace{-1em}
	\caption{\label{fig:bugpattern_02}Example Query In Loop bug pattern.}
	\vspace{-1em}
\end{figure}

\paragraph{Query In Loop:}
Physical buttons or sensors on a robot are not only activated for an instant when they are triggered. Instead, buttons remain pressed and sensors return values for a period of time.
When a query is used as a condition of a fast loop (i.e., a loop without time or wait statements), the query is repeated very frequently, in some cases even several hundred times per second. Thus, the code reacting to detected values is executed uncontrollably often and can falsify other values or states that are changed according to the query result. 
To prevent this, sensor queries within a loop that change values should be avoided or at least secured by waiting statements. \Cref{fig:bugpattern_02} shows an example of this bug pattern.

\begin{figure}[t]
	\centering
	\includegraphics[scale=0.25]{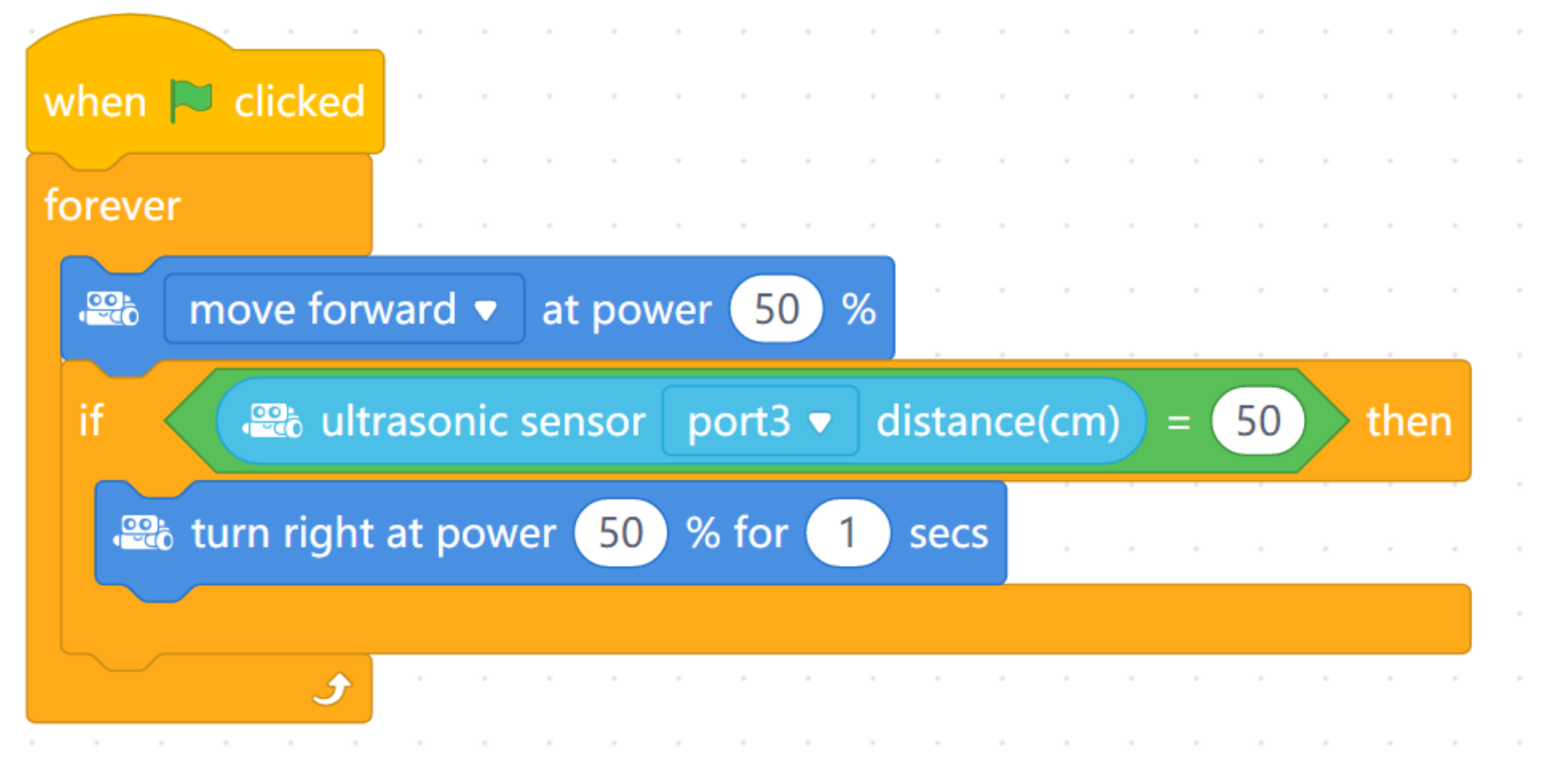}
	\vspace{-1em}
	\caption{\label{fig:bugpattern_11}Example Sensor Equals Check bug pattern.}
\end{figure}

\paragraph{Sensor Equals Check:}
The values of the robots' sensors are not rounded and are therefore rarely at a single exact value.
If a sensor is compared to an exact value in the code, the probability of this state never occurring is very high. 
Therefore, when querying the sensors, one should compare to a range of values instead of exact values. An example of this bug pattern is shown in \Cref{fig:bugpattern_11}. The line-following sensor is an exception, as the values for this sensor are just integers from 0 to 3, allowing an exact comparison. 

\paragraph{Several Launches:} 
When used with the `upload mode', the \mbot robot cannot execute scripts in parallel, but only sequentially. In addition, the board launch event is the only event that works in upload mode.
If two board launch scripts are programmed in parallel, they are only executed one after the other, or --- in the case of forever loops --- sometimes not at all. 
For parallel programming, either the `live mode' or a \codeyrocky robot must be used. 

\paragraph{Stuttering Action:}
In live mode, the calculation of all scripts is taken over by the computer, and the corresponding commands are sent to the robot individually. This runs noticeably slower than when code is uploaded to the robot. 
If individual time-limited statements are in a loop in live mode, they are not executed smoothly one after the other. Instead, the code stops briefly between each block. 
To prevent this, time-limited statements in loops should generally be avoided and replaced by constant move blocks guarded by a query in order to have a stopping condition.

\paragraph{Useless Sensing:} If the value range of a query involving a specific type of sensor is out of the allowed range, then coresponding conditions will either always be true, or always be false. We distinguish several variants of this pattern depending on the sensor involved:
\begin{itemize}
\item \textbf{Useless Battery Sensing:} The value range for Battery Sensing is from 0 to 100. 

\item \textbf{Useless Colour Sensing:} 
The value range for Colour Sensing is from 0 to 255.

\item \textbf{Useless Distance Sensing:} 
The value range for Distance Sensing is from 3 to 400.

\item \textbf{Useless Light Sensing:} 
The values for Light Sensing is from 0 to 100 for the \codeyrocky and from 0 to 1020 for the \mbot.

\item \textbf{Useless Line Sensing:} 
The values for Line Sensing are the integers from 0 to 3.

\item \textbf{Useless Loudness Sensing:} 
The value range for Loudness Sensing is from 0 to 100.

\item \textbf{Useless Pitch Angle Sensing:} 
The value range for Pitch Angle Sensing is from -180 to 180.

\item \textbf{Useless Potentiometer Sensing:} 
The value range for Potentiometer Sensing is from 0 to 100.

\item \textbf{Useless Roll Angle Sensing:} 
The value range for Roll Angle Sensing is from -90 to 90.

\item \textbf{Useless Shaking Sensing:} 
The value range for Shaking Sensing is from 0 to 100.
\end{itemize}

\begin{figure}[t]
	\centering
	\includegraphics[scale=0.3]{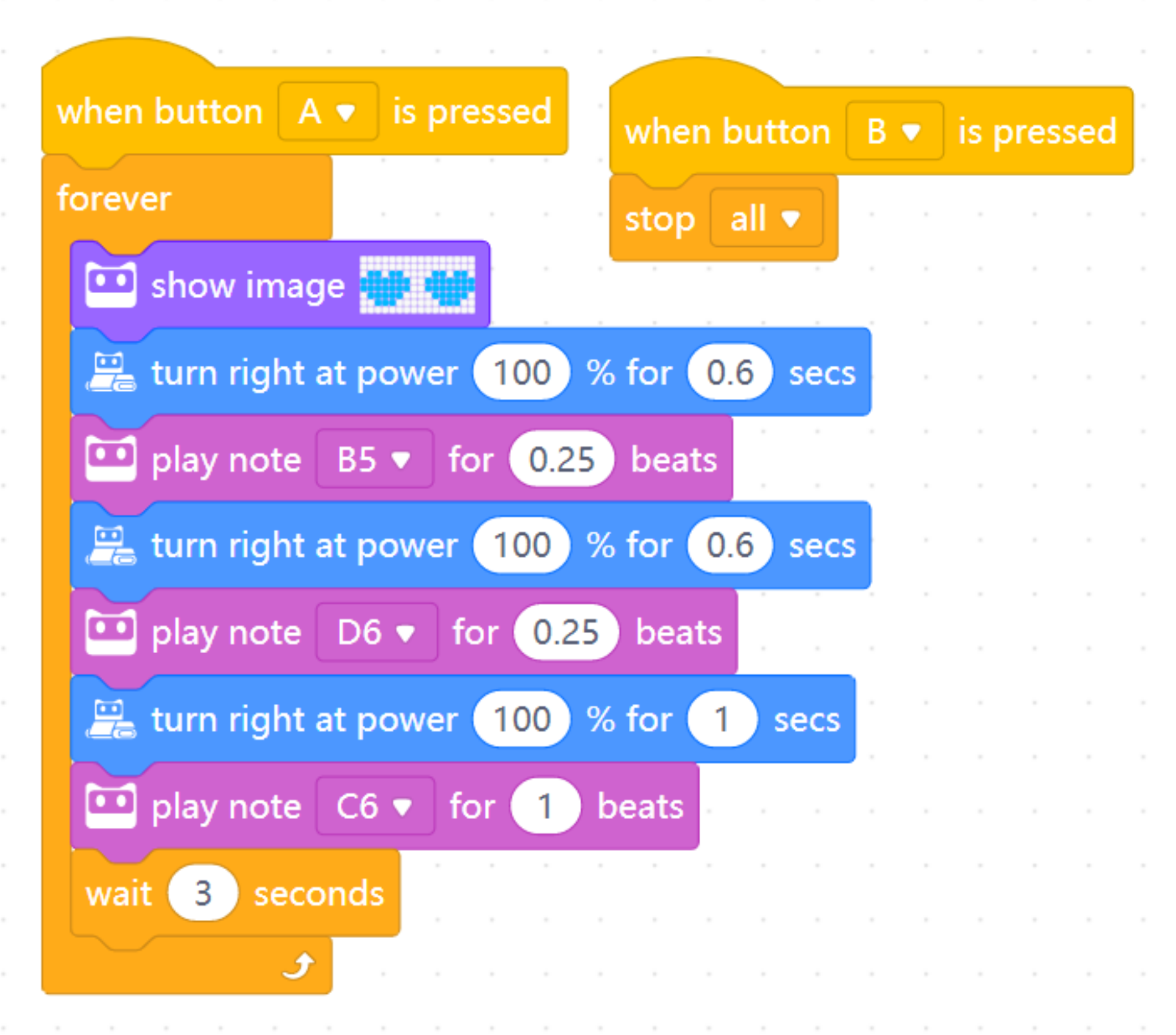}
	\vspace{-1em}
	\caption{\label{fig:bugpattern_01}Example Waiting Aborted bug pattern.}
\end{figure}

\paragraph{Waiting Aborted:} 
\looseness=-1 
\mblock offers blocks with a time limit, which activate a specific actuator for
the specified time. When a program with such blocks is uploaded to the robot,
the time-limited statement is split into three separate parts: activation,
waiting, and deactivation. Executing a parallel script which cancels all
scripts on a \codeyrocky robot can lead to the following problem: When the
scripts are cancelled during the waiting phase of the time-limited statement,
the waiting itself is cancelled, but also, the deactivation will no longer be
carried out. This can lead to motors running indefinitely or lights not being
switched off. The bug can be prevented by using a detailed stop script that
stops all affected actuators before the scripts are terminated.
\Cref{fig:bugpattern_01} shows an example of this bug pattern.

\subsection{Code Smells}
A code smell in \mblock is a code idiom that increases the probability of errors in a program or decreases the readability of the code by using unexpected values when addressing sensors and actuators.

\paragraph{Negative Motor Power:}
Using negative motor power values is possible and leads to the robot changing its direction. I.e., when using negative values for moving forward, the robot moves backward, and vice versa. The same applies when turning left or right. Since using negative values reduces the code readability, one should rather use the corresponding blocks (e.g., move backward with positive value instead of move forward with negative value).

\paragraph{Non-effective Modification:} Changing the settings of a robot attribute several times in a row can lead to the robot behaving differently than expected: Without waiting phases in between the modifications, the attributes will be set to the next value so fast that one cannot see any effect. Only the last setting will be visible. In order to notice every single modification, time-limited statements or waiting blocks in between the attribute settings have to be used.

\looseness=-1 
\paragraph{Non-effective Time Limit:} When the time value of a time-limited block is set to 0, the execution of the block will have no effect. Writing unnecessary code is bad for readability and should be avoided.

\subsection{Code Perfumes}
Code perfumes for \mblock are
aspects of code that correctly apply concepts related to robot use, in
particular correct use of sensory data and actuators.

\paragraph{Colour Usage:}
Appropriate values for colours are integers from 0 to 255.

\looseness=-1 
\paragraph{Correct Sensing:} Queries involving a sensor should use a valid value range. This indicates comprehension of the sensing concept, which is frequently used in robot programming. Depending on the sensor used, we distinguish several variants of this code perfume:
\begin{itemize}
	\item \textbf{Battery Sensing:}
	The value range for Battery Sensing is from 0 to 100. 
	
	\item \textbf{Colour Sensing:}
	The value range for Colour Sensing is from 0 to 255.
	
	\item \textbf{Distance Sensing:}
	The value range for Distance Sensing is from 3 to 400.
	
	\item \textbf{Light Sensing:}
	The value range for Light Sensing is from 0 to 100 on the \codeyrocky and from 0 to 1020 on the \mbot.
	
	\item \textbf{Line Sensing:}
	The values for Line Sensing are the integers from 0 to 3.
	
	\item \textbf{Loudness Sensing:}
	The value range for Loudness Sensing is from 0 to 100.
	
	\item \textbf{Pitch Angle Sensing:}
	The value range for Pitch Angle Sensing is from -180 to 180.
	
	\item \textbf{Potentiometer Sensing:}
	The value range for Potentiometer Sensing is from 0 to 100.
	
	\item \textbf{Roll Angle Sensing:}
	The value range for Roll Angle Sensing is from -90 to 90.
	
	\item \textbf{Shaking Sensing:}
	The value range for Shaking Sensing is from 0 to 100.
\end{itemize}

\paragraph{Correct Actuator Deactivation:}
When using blocks that activate an actuator, one must be aware of the necessity of also deactivating them. Writing a separate script for turning the actuators off is often useful and shows that this robot specific usage of actuators has been understood. We define several versions of this code perfume depending on the specific actuator used:
\begin{itemize}
	\item \textbf{LED Off}	
	\item \textbf{Light Off}
	\item \textbf{Matrix Off}
	\item \textbf{Motor Off}
\end{itemize}

\looseness=-1 
\paragraph{Loop Sensing:}
In order to make the robot react to a specific change of a sensor's value, one must continuously read the corresponding sensor values. Using queries concerning sensors within a loop indicates the comprehension of this robot typical concept of sensing.

\paragraph{Motor Usage:}
The motors of the robots can be controlled with a minimum of 0\% and a maximum of 100\%. When using the \mbot robot, values beneath 25\% have the same effect as 0\%. Therefore, appropriate values range from 0 to 100 for the \codeyrocky robot, and from 25 to 100 for the \mbot robot.

\paragraph{Parallelisation:}
Writing several scripts with the same hat block can be indicative of attempts to implement independent
subtasks at a higher readability level.

\section{Experimental Setup}
\label{sec:setup}
\looseness=-1
To evaluate the different types of code patterns for \mblock projects,
 we empirically investigated the following research questions:
\begin{itemize}
	\item \textbf{RQ1:} How does \mblock code compare to \Scratch code?
	\item \textbf{RQ2:} How common are robot bug patterns, code smells and code perfumes in \mblock programs?
	\item \textbf{RQ3:} How severe are the bug patterns found?
\end{itemize}

\subsection{Analysis Tool}
\looseness=-1
In order to study the occurrence of the patterns listed in \cref{sec:patterns} in
\mblock programs, we extended the \litterbox~\cite{fraser2021} tool, which was
originally intended for \Scratch projects (see \cref{sec:background}). \litterbox handles the analysis of \Scratch programs by automatically converting a project
into an abstract syntax tree (AST), and then checking for the presence of block
combinations utilising a visitor pattern. Each bug pattern, code perfume and code smell has its own visitor looking for the block combination defining the code pattern.
Since \mblock is a fork of \Scratch
adding robot functionalities, the extension of \litterbox required us to
implement handling for all \mblock blocks in the parser, so that programs for
\codeyrocky and \mbot can be represented as ASTs (which then consist of both
standard \Scratch nodes as well as \mblock-specific nodes). For each of the finders described in \cref{sec:patterns} we then implemented an
AST visitor that traverses the blocks of a program and reports all matches
found. Besides this AST visitor, each pattern also consists of a textual hint
defined in multiple languages, which can be shown to a user checking their
program for patterns.

\subsection{Dataset}

\subsubsection{\mblock dataset}
\looseness=-1
As subjects of our study we created a data set of 28,192 \mblock programs by
mining all publicly shared projects from the \mblock
website\footnote{\url{https://planet.mblock.cc/}, last accessed 01.06.22} until
the first quarter of 2021. Out of all these projects, 329 resulted in parsing
exceptions when attempting to apply \litterbox. Of these, 32 were empty
files without any code, while the remaining 297 projects contain robot features
not supported by our extension (which focuses on \codeyrocky and \mbot
features). This leaves a data set of 27,863 programs for analysis.

\looseness=-1
\Scratch programs generally organise code in terms of different actors
(i.e., sprites, background). \mblock programs add dedicated actors for robots, so we can identify which type of robot an \mblock program is intended for by
checking which actors exist. Out of the 27,863 programs, 16,569 contain a
\codeyrocky or \mbot robot (and sometimes more than one). The remaining 11,294
programs either contain code for other robots, or mostly are regular \Scratch
programs, since \mblock can also be used as a \Scratch programming environment. Since the \mblock programming environment used to add a \codeyrocky robot actor by
default in the past, we further filtered the dataset by removing all programs where the
\mbot and \codeyrocky actors contain no code. This leaves a final dataset of
3,540 relevant projects with a total of \numCodeyProj \codeyrocky robots and \numMcoreProj \mbot
robots, including \countbothrobots projects utilising both robots. For the remainder of this paper, the 3,540 \mblock projects are used as the robot
dataset.

\subsubsection{\Scratch dataset}
\looseness=-1
In order to compare robot code with regular \Scratch code, we created a
comparable set of \Scratch programs in addition to the existing robot set. We
used the \Scratch REST-API to mine a dataset of more than 1 million projects,
of which we randomly sampled 3,540 non-empty, non-remixed programs.

\subsection{Methodology}

\subsubsection{RQ1: \mblock vs. \Scratch differences}
\looseness=-1
We applied \litterbox on the \Scratch and the \mblock dataset and collected all
information about bug patterns, code smells, code perfumes, and code metrics
available in \litterbox, including those added by our \mblock extension. To answer RQ1 we
characterise the differences between regular \Scratch programs and \mblock
programs in terms of the code metrics, compare the programs in terms of
traditional patterns found, as well as overall number of findings reported by
\litterbox. For all significance tests we used a non-parametric test, the Mann-Whitney-U test~\cite{mann1947}, with $\alpha = 0.05$, as this test is designed for independent samples.
The effect sizes are calculated using Vargha-Delaney's $\hat{A}_{12}$~\cite{vargha2000}.
In our context, the $\hat{A}_{12}$ is an estimation of the probability
that, if we extract a metric using \litterbox on an \mblock program,
we will obtain higher values than extracting the metric on a \Scratch
program. If \mblock and \Scratch programs are equivalent with respect
to a metric, then $\hat{A}_{12} = 0.5$. A high value
$\hat{A}_{12} = 1$ means that the metric is higher on all \mblock
programs, a low value $\hat{A}_{12} = 0$ means that the metric is
higher on all \Scratch programs.


\subsubsection{RQ2: \mblock pattern occurrences}
\looseness=-1
To answer this research question, we consider the findings reported for all the
 \mblock-specific patterns on the \mblock
dataset. For each pattern, we inspect the total number of instances found, how
many programs are affected, and how the patterns relate to program complexity,
as measured by different metrics such as the weighted method count (i.e., sum
of cyclomatic complexities of all scripts).

\subsubsection{RQ3: \mblock bug pattern severity}
\looseness=-1
To answer RQ3, we specifically consider the \mblock bug patterns. For each bug pattern, we randomly selected 10 projects that contain at least one such bug, or
all bugs if there are less than 10 projects with that bug pattern overall.
Then, two authors manually classified the projects into the three categories
\emph{failures} (the bug pattern causes incorrect program behaviour), \emph{not
executed} (the bug pattern may represent incorrect code, but this code cannot
be executed), and \emph{false positives} (the bug pattern does not affect the
execution). The classification requires executing the programs using a robot.
In case of disagreement between the two raters, the cases were discussed among
the authors until a consensus was found.

\subsection{Threats to Validity}
\looseness=-1
\noindent\emph{External validity}: Although we used a large dataset
of \mblock programs, and an equally large set of \Scratch programs, results may
not generalise to other programs. In particular, our data mining approach can
only download publicly shared projects, and incomplete programs may not be
shared.

\looseness=-1
\noindent\emph{Internal validity}: We thoroughly tested our
implementation to avoid bugs in the implementation. To reduce the threat of
misclassification for RQ3, two authors independently classified findings.

\looseness=-1
\noindent\emph{Construct validity}: While we measured the frequency
of all patterns, we only evaluated severity for bug patterns, and generally did
not evaluate their effects on learners.

\section{Results}\label{sec:evaluation}

\subsection{RQ1: How Does \mblock Code Compare to \Scratch Code?}

\subsubsection{Size and Complexity}
\begin{figure}[t]
	\centering
	\includegraphics[width=\columnwidth]{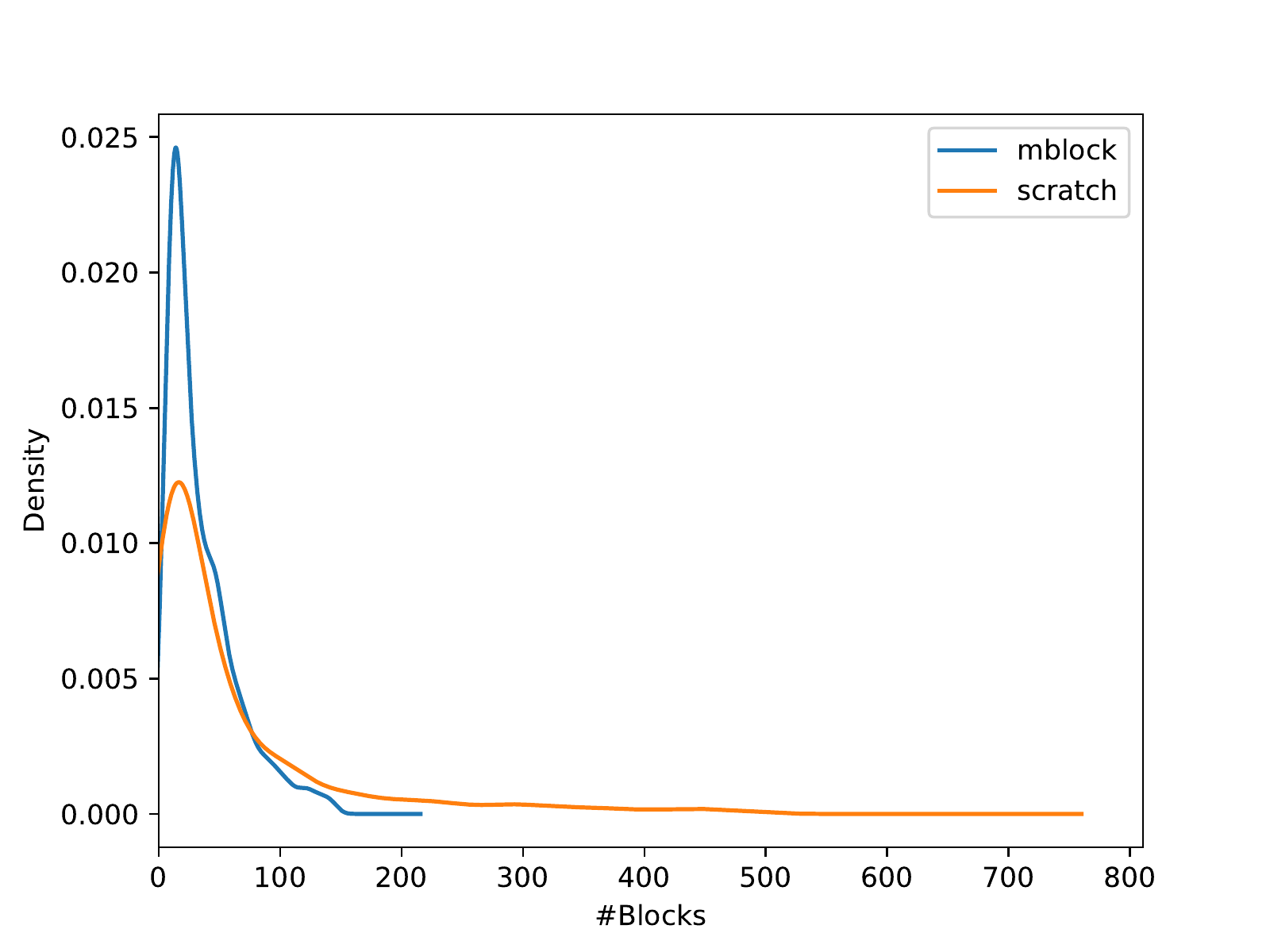}
	\vspace{-1em}
	\caption{\label{fig:blocks_density}Density of the block count compared in \Scratch and \mblock programs.}
\end{figure}

\Cref{tab:metrics_table} compares the programs in our \mblock and \Scratch
datasets in terms of size and complexity metrics: On average, \Scratch projects
have substantially and significantly more blocks than \mblock programs. This is
also confirmed by the distribution of sizes shown in \cref{fig:blocks_density},
with \Scratch programs having up to \maxblocksscratch blocks, while no \mblock program is
larger than \maxblocksmblock blocks. The overall larger complexity is also confirmed by the
mean weighted method count (WMC), which again is significantly higher for
\Scratch programs (average of 29.88) compared to \mblock (8.97 on average).
This may be attributed to the limited scope of robot programs: There is only so much 
you can do with the same set of sensors and actuators, while the boundary of possibilities 
in \Scratch will not be reached so soon and students may create more complex programs.

One factor leading to fewer scripts in robot programs may be that \mbot only
allows one script when working in upload mode as described in the
\textit{Several Launches} bug pattern. This is confirmed by the statistically
significant difference in the number of scripts in \cref{tab:metrics_table} and when comparing to \mbot programs only.
While we also observe a statistically significant difference in the length of
the longest script (\cref{tab:metrics_table}), the cyclomatic complexity of the
most complex script per project is \emph{not} significantly different
($p=\pvaluemostcomplexscript$). Consequently, it seems that the more complex
nature of \Scratch projects lies in the use of more parallelism, while
individual scripts can be just as complex in \mblock.

\subsubsection{Bug Patterns} An alternative way to understand the differences
between the programs is by considering how many general \Scratch patterns
(i.e., excluding the new \mblock-patterns) are found in the different types of
programs. In the total of 3,540 programs each, \classicbugpatternmblock bug
patterns were found for the \mblock projects and \classicbugpatternscratch for
the \Scratch dataset, which is a large and statistically significant difference
($p\pvalueclassicbugpattern$, $\hat{A}_{12}=\aclassicbugpattern$). Since \Scratch programs
contain more blocks, there naturally are more possibilities for bug patterns,
but this difference remains significant even when normalising by the number of
blocks in a program ($p\pvalueclassicbupgatternperblockcount$,
$\hat{A}_{12}=\aclassicbupgatternperblockcount$).
On the one hand, this suggests that \mblock programs only use a subset of the
functionalities of \Scratch: For example, some \Scratch bug patterns (e.g.,
\emph{Message Never Received}, \emph{Forever Inside If})~\cite{fraser2021} occur also in \mblock programs,
while aspects like cloning of sprites generally are not contained in \mblock
programs. On the other hand, \mblock programs add new functionality that may
not be captured by the existing bug patterns. When considering the total number
of bug patterns, including the \mblock-specific ones introduced in this paper,
the difference between the datasets is no longer significant
(p=\pvalueallbugpattern). The total numbers are shown in \cref{tab:pattern_table}.
This confirms the need for robot-specific bug
patterns, and demonstrates that our patterns capture the robot
functionality well.

\subsubsection{Code Smells} The comparison in terms of \Scratch code smells paints
a similar picture: There are \classicsmellmblock in the \mblock projects and
\classicsmellscratch for \Scratch, which is a significant difference
($p\pvalueclassicsmell$, $\hat{A}_{12}=\aclassicsmell$) even when normalising for block count
($p=\pvalueclassicsmellperblockcount$, $\hat{A}_{12}=\aclassicsmellperblockcount$). Unlike for
bug patterns, however, even when considering also the \mblock-specific code
smells, the difference remains significant ($p\pvalueallsmell$, $\hat{A}_{12}=\aallsmell$).
This is because code smells tend to be more concerned with the general
structure of programs than with specific functionality. While code smells like
\emph{Code Clones}, \emph{Duplicated Script} and \emph{Long Script}~\cite{fraser2021} also occur in \mblock programs,
the simpler structure of \mblock programs provides less opportunities for such smells. A noteworthy exception for \mblock is \emph{Empty Sprite},
since \mblock initialises an empty graphical sprite by default, even though it
is not needed for robots.

\subsubsection{Code Perfumes} Code perfumes are similarly imbalanced, with a total
of \classicperfumemblock \Scratch perfumes for \mblock programs, and
\classicperfumescratch for \Scratch programs ($p\pvalueclassicperfume$,
$\hat{A}_{12}=\aclassicperfume$, and $p\pvalueclassicperfumeperblockcount$,
$\hat{A}_{12}=\aclassicperfumeperblockcount$ when normalising by number of blocks). Like
code smells, many code perfumes are concerned with control flow aspects that
seem to occur less frequently in \mblock programs, as also suggested by the
higher complexity of \Scratch programs (\cref{tab:metrics_table}). The
imbalance remains even when including all new code perfumes
($p\pvalueallperfume$, $\hat{A}_{12}=\aallperfume$). This is because the \mblock specific
perfumes like the four \textit{Off} variants are only needed once per project
and thus do not increase the count of perfumes by much. Furthermore, a \Scratch
project with multiple sprites may need multiple collision checks, whereas \codeyrocky and \mbot programs usually just need a single check for obstacles ahead.

%

\begin{table}
\centering
\caption{Mean values of metrics for \Scratch and \mblock datasets compared.}
\vspace{-1em}
\label{tab:metrics_table}
\begin{tabular}{l@{}rrrr}
\toprule
{} & \mblock & \Scratch & $p$-value &   $\hat{A}_{12}$ \\
\midrule
\#Blocks             &   51.25 &   127.48 &  <0.001 &  0.47 \\
WMC                 &    8.97 &    29.88 &  <0.001 &  0.38 \\
\#Scripts            &    3.89 &    13.84 &  <0.001 &  0.32 \\
Longest Script      &   15.52 &    16.38 &  <0.001 &  0.53 \\
Most Complex Script &    3.54 &     4.18 &     0.4 &  0.51 \\
\bottomrule
\end{tabular}
\end{table}

\begin{table}
	\centering
	\caption{Number of pattern instances found using only the \Scratch finders or all incl. \mblock on both the \Scratch and \mblock datasets.}
	\vspace{-1em}
	\label{tab:pattern_table}
	\begin{tabular}{lrrrr}
		\toprule
		{} & \multicolumn{2}{c}{\Scratch} &\multicolumn{2}{c}{ All} \\
		Patterns & \mblock & \Scratch & \mblock & \Scratch \\
		\midrule
		Bug patterns &  \classicbugpatternmblock   &   \classicbugpatternscratch  & \allbugpatternmblock & \allbugpatternscratch \\
		Code smells &        \classicsmellmblock &     \classicsmellscratch & \allsmellmblock & \allsmellscratch \\
		Code perfumes &     \classicperfumemblock &     \classicperfumescratch & \allperfumemblock & \allperfumescratch \\
		\bottomrule
	\end{tabular}
\end{table}

\summary{RQ1}{\Scratch and \mblock programs are significantly different regarding their size and complexity. Robot-specific patterns are needed to analyse \mblock programs as thoroughly as \Scratch projects.}

\subsection{RQ2: How Common are Robot Bug Patterns, Code Smells and Code Perfumes in \mblock Programs?}

\subsubsection{Bug Patterns} 

We found instances for \numbugpatternsfound of the \numbugpatterns robot bug patterns in the dataset of random
\mblock projects. In total, there are \numBugpattern bug pattern instances, and
\numprojectsBugpattern projects contain at least one bug pattern.
\Cref{tab:bugpattern_table} summarises the number of bug pattern instances
found for each type, the number of projects containing at least one instance of
the respective pattern, and the average weighted method count of these
projects. Note that one project may contain more than one type of bug pattern.

\begin{table}
\centering
\caption{Number of bug pattern instances found in total and number of projects containing the bug pattern.}
\label{tab:bugpattern_table}
\vspace{-1em}
\resizebox{\columnwidth}{!}{%
\begin{tabular}{l@{}rrr}
\toprule
                   Bug Pattern & \# Patterns & \# Projects &  Avg. WMC \\
\midrule
           Action Not Stopped &        117 &         77 &     34.34 \\
          Colour Out Of Range &          9 &          7 &      4.86 \\
     Interrupted Loop Sensing &      1,287 &        319 &     14.34 \\
              LED Off Missing &      1,516 &        675 &     11.37 \\
            Light Off Missing &         12 &          9 &      5.56 \\
              Low Motor Power &        348 &         90 &     11.38 \\
           Matrix Off Missing &        694 &        478 &     13.34 \\
         Missing Loop Sensing &        278 &        126 &     12.20 \\
            Motor Off Missing &        491 &        364 &     11.55 \\
           Motor Out Of Range &        230 &         68 &      9.53 \\
        Parallel Actuator Use &        887 &        473 &     16.21 \\
                Query In Loop &         26 &          8 &     49.88 \\
          Sensor Equals Check &         70 &         34 &     17.71 \\
             Several Launches &         52 &         52 &     18.85 \\
            Stuttering Action &         89 &         52 &     13.02 \\
      Useless Battery Sensing &          0 &          0 &           \\
       Useless Colour Sensing &          0 &          0 &           \\
     Useless Distance Sensing &         10 &          7 &      8.29 \\
        Useless Light Sensing &          1 &          1 &     12.00 \\
         Useless Line Sensing &          3 &          3 &     12.00 \\
     Useless Loudness Sensing &          0 &          0 &           \\
  Useless Pitch Angle Sensing &          0 &          0 &           \\
Useless Potentiometer Sensing &          0 &          0 &           \\
   Useless Roll Angle Sensing &          0 &          0 &           \\
      Useless Shaking Sensing &          0 &          0 &           \\
              Waiting Aborted &          9 &          9 &     10.11 \\
\midrule
                        Total &      6,129 &      1,903 &     11.39 \\
\bottomrule
\end{tabular}%
}
\end{table}

The bug patterns that were not found in the dataset are all related to sensors
that are only found on the \codeyrocky, for which the dataset only contains
\numCodeyProj projects. Even within these projects, these sensors are only
rarely used, as shown in \cref{tab:sensor_table}: These sensors were mostly used
for calculations or to show the value on the LED Matrix, but not in a way that
the finders would check. However, since we found similar bug patterns for other sensors
like \emph{Useless Distance Sensing}, we expect that a growing
community of \codeyrocky users will lead to all sensors being used in the
future.

\begin{table}
\centering
\caption{Number of projects using a specific sensor.}
\label{tab:sensor_table}
\vspace{-1em}
\begin{tabular}{lr}
\toprule
          Sensor & \# Projects \\
\midrule
   Battery Level &         11 \\
Colour Detection &          3 \\
            Gyro &          4 \\
 Light Intensity &        292 \\
  Line Following &        225 \\
        Loudness &         11 \\
   Potentiometer &         20 \\
         Shaking &          5 \\
     Ultra Sonic &        736 \\
\bottomrule
\end{tabular}
\end{table}

The most frequent bug pattern is \emph{LED Off Missing} with 1,516 instances.
Combined with the frequent other \emph{Off Missing} bug patterns, it appears
that beginners do not pay attention to returning the robot to a neutral state when the
program has finished execution. Note that \emph{Light Off Missing} is an
exception here with only 12 instances found, but again this may be caused by
the low number of \codeyrocky projects.
This conjecture is further supported when considering the number of projects
containing these bug patterns (Column 3 in \cref{tab:bugpattern_table}): Here
\emph{LED Off Missing} is also ranked first (675) followed by \emph{Matrix Off
Missing} (478) and \emph{Parallel Actuator Use} (473). Furthermore, the
remaining non \codeyrocky exclusive bug pattern concerning the switching off of
actuators (i.e., \emph{Motor Off Missing}) come in at fourth place with 364
projects exhibiting it.

The second and third most frequent bug patterns are \emph{Interrupted Loop
Sensing} (1,287) and \emph{Parallel Actuator Use} (887). Both of these can come from a wide range of blocks, and the causes --- either using a timed block
that interrupts a sensing process or an accidental use of the same actuator at the
same time --- are easy to create.

The least common bug patterns (except for those not occurring at all) are
\emph{Useless Light Sensing} (1) and \emph{Useless Line Sensing} (3).
The Light Intensity and Line Following sensors are rather easy to use,
as Light Intensity has the biggest range of all sensors and Line Following
uses just four integers instead of a range. Consequently, it is more
difficult to make mistake here.

The average complexity of the projects containing bug patterns (column WMC in
\cref{tab:bugpattern_table}) reveals that the least complex projects are found
for \emph{Colour Out of Range} (4.86), \emph{Light Off Missing} (5.56),
\emph{Useless Distance Sensing} (8.29) and \emph{Motor Out of Range} (9.53).
Out of these, \emph{Colour Out of Range}, \emph{Light Off Missing} and
\emph{Motor Out of Range} relate to very basic behaviour that can be
implemented even in very small projects at early stages of programming.
%

The bug patterns appearing in the most complex projects are \emph{Query in
Loop} (49.88) and \emph{Action Not Stopped} (34.34). 
\emph{Query in Loop} requires variables, which are a rather
advanced concept and are not as frequently used in \mblock as they are in
text-based programming languages. The same holds true for \emph{Action Not
Stopped}, as this pattern needs multiple scripts and control structures, which
themselves increase the WMC.

\subsubsection{Code Smells} 

We found instances for all code smells in the dataset. In total, there are
\numSmell code smell instances, and \numprojectsSmell programs contained at
least one code smell. \Cref{tab:smell_table} shows the numbers of code smell
instances found for each type, the number of projects containing at least one
instance of the respective code smell, and the average weighted method count of
these projects. Again one project may contain more than one type of code smell.

\begin{table}
\centering
\caption{Number of code smell instances found in total and number of projects containing the code smell.}
\label{tab:smell_table}
\vspace{-1em}
\begin{tabular}{l@{}rrr}
\toprule
                 Code Smell & \# Patterns & \# Projects &  Avg. WMC \\
\midrule
Negative Motor Power &        257 &         71 &     11.42 \\
Non-effective Modification &        323 &         86 &     14.22 \\
Non-effective Time Limit &         12 &          8 &      5.38 \\
\midrule
                 Total &        592 &        157 &     11.34 \\
\bottomrule
\end{tabular}
\end{table}

For all three aspects (i.e., number of pattern instances, number of projects
showing the pattern and average WMC) the ranking is the same, with
\emph{Non-effective Modification} followed by \emph{Negative Motor Power} and
\emph{Non-effective Time Limit}. \emph{Non-effective Modification} has a wide range of
blocks that can cause the smelly situation. 
The large number of \emph{Negative Motor Power} code smells may be caused by copy\&paste, as it may be quicker to copy a forward block and enter a negative number, rather than looking for the backward block.

The small number of \emph{Non-effective Time Limit} code smells is likely explained
by useless blocks quickly cluttering the program, so that they are frequently
removed. The low average WMC for this code smell also suggests that the
projects using blocks without effect are rather simple and may be from
beginners in their first projects trying different values and experimenting
with blocks.

\subsubsection{Code Perfumes} 

We found instances for \numperfumesfound of the \numperfumes code perfumes defined in
\cref{sec:patterns}. In total there are \numPerfumes code perfume instances,
and \numprojectsPerfumes projects containing at least one code perfume.
Analogous to the other code patterns, \cref{tab:perfumes_table} shows the
number of code perfume instances found for each type, the number of projects
containing at least one instance of the respective code perfume, and the average
weighted method count of these programs. Once again, a project may contain more
than one type of code perfume.

\begin{table}
\centering
\caption{Number of code perfume instances found in total and number of projects containing the code perfume.}
\label{tab:perfumes_table}
\vspace{-1em}
\begin{tabular}{l@{}rrr}
\toprule
             Code Perfume & \# Patterns & \# Projects &  Avg. WMC \\
\midrule
      Battery Sensing &          1 &          1 &      9.00 \\
       Colour Sensing &          0 &          0 &           \\
         Colour Usage &        829 &        291 &     12.63 \\
     Distance Sensing &      1,204 &        548 &     11.90 \\
              LED Off &          9 &          9 &      7.56 \\
            Light Off &          0 &          0 &           \\
        Light Sensing &        259 &        178 &     12.89 \\
         Line Sensing &        584 &        138 &     25.72 \\
         Loop Sensing &        573 &        272 &      9.36 \\
     Loudness Sensing &         12 &          8 &     13.75 \\
           Matrix Off &          2 &          2 &      9.50 \\
            Motor Off &         43 &         43 &      9.19 \\
          Motor Usage &     10,896 &      1,815 &      9.64 \\
      Parallelisation &         72 &         40 &     14.65 \\
  Pitch Angle Sensing &          1 &          1 &     40.00 \\
Potentiometer Sensing &          5 &          3 &     11.67 \\
   Roll Angle Sensing &          2 &          1 &    102.00 \\
      Shaking Sensing &          3 &          3 &     22.00 \\
\midrule
                Total &     14,495 &      2,284 &      9.44 \\
\bottomrule
\end{tabular}
\end{table}

Two code perfumes were not found in the dataset: The \emph{Colour Sensing} code
perfume depends on the Colour Detection sensor, which is rarely used (see
\cref{tab:sensor_table}). For \emph{Light Off} the corresponding bug pattern
was also rare, and the likely reason is that there are only \numCodeyProj
\codeyrocky projects.

The most frequent code perfume is \emph{Motor Usage} with 10,896 instances. It
is followed by \emph{Distance Sensing} (1,204) and \emph{Colour Usage} (829).
Both perfumes concerning the correct usage of actuators in the top three
represent the easiest and most basic way of working with robots. \emph{Distance
Sensing} is also to be expected as it relates to the most used sensor. The
ranking is the same when considering the number of projects in
\cref{tab:sensor_table}.

\emph{Battery Sensing} and \emph{Pitch Angle
Sensing} occurred only once each. These two perfumes, as well as other
related code perfumes such as \emph{Roll Angle Sensing} (2),
\emph{Shaking Sensing} (3) and \emph{Potentiometer Sensing} (5), all are based on
sensors which are not used frequently in our dataset (see
\cref{tab:sensor_table}).
\emph{Matrix Off} (2) and and \emph{LED Off} (9) are also relatively rare; the
bug patterns which are the counter pieces to these two code perfumes are among
the most frequent bug patterns, demonstrating that returning the robot to a
neutral state after program execution is not frequently done.


Considering project complexity (\cref{tab:perfumes_table}), \emph{Roll Angle
Sensing} (102.00) and \emph{Pitch Angle Sensing} (40.00) are contained in only one project each, which happen to be more complex projects.
The fact that \emph{Line Sensing} is used in more complex projects (25.72)
seems surprising at first, as it is a common and easy task. However, line
following tasks tend to require several control structures, i.e., at least one
loop and then one if block for each of the four states of the sensor; this
explains the higher complexity.

The average complexity of projects containing \emph{Loop Sensing} (9.36) is low because most robot programs require some sensing loop in order to react to influences from the real world. The low complexity of projects containing \emph{LED Off}, \emph{Motor Off}, and \emph{Matrix Off} shows that it is not difficult to correctly turn off actuators. We conjecture that actuators are rarely turned off not because it is difficult, but because users mostly are not aware it should be done.

\summary{RQ2}{Bug patterns, code smells and code perfumes for robot projects appear frequently in \mblock programs regardless their complexity.}

\subsection{RQ3: How Severe are the Bug Patterns Found?}

\begin{figure}[t]
	\centering
	\includegraphics[width=\columnwidth]{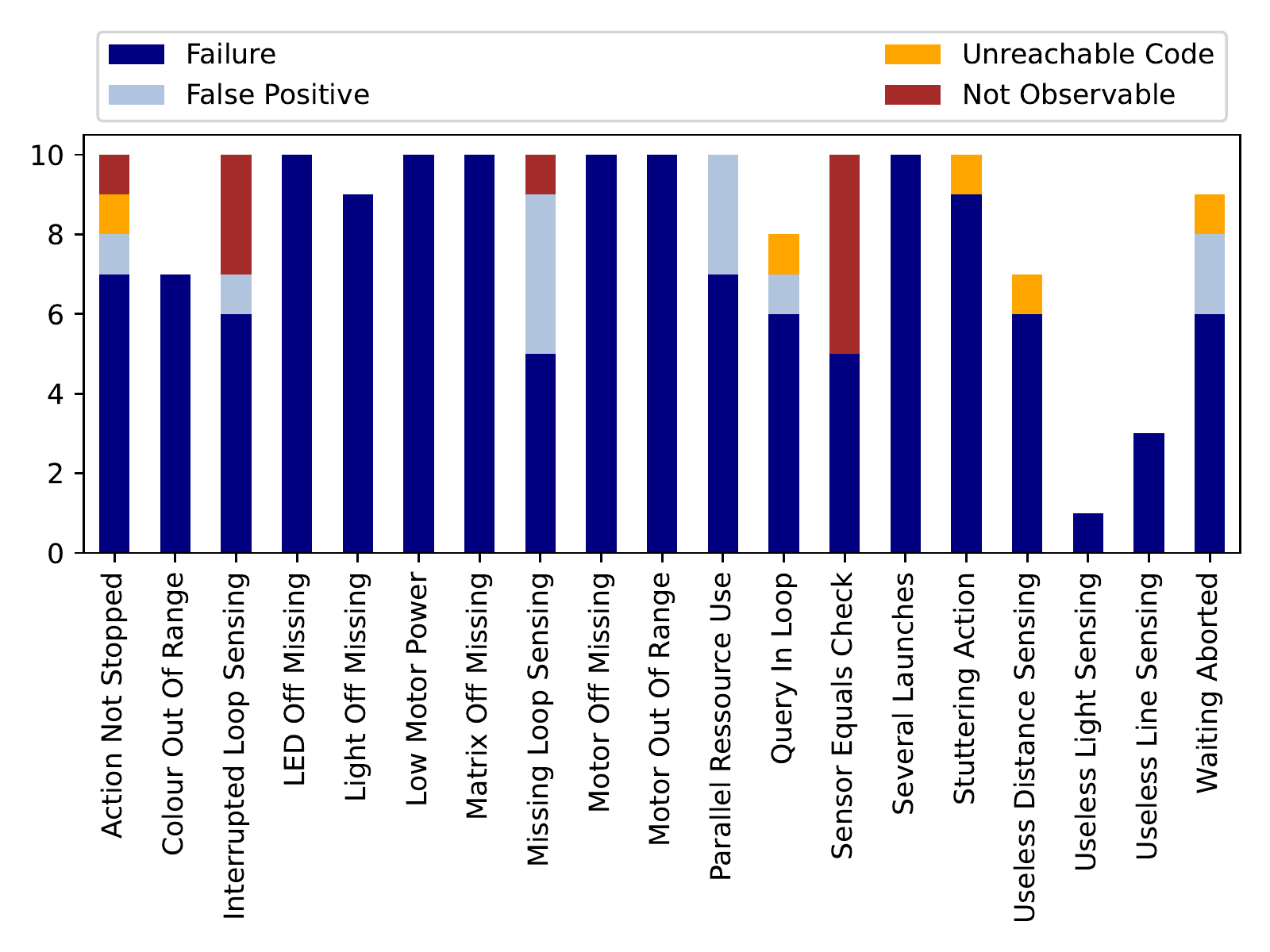}
	\vspace{-1em}
	\caption{\label{fig:classification}Results of the manual classification.}
\end{figure}

Figure~\ref{fig:classification} shows the results of the manual classification
for each of the \numbugpatternsfound bug patterns with at least one occurrence
in the dataset. Out of the \numclassified inspected projects, \numfailure
instances of bug patterns manifested into failures, \numunreachable did not
result in failures because the defective code was never executed,
\numnotnoticable were defects without visible impact, and \numfalsepositive
bugs were classified as false positives.

Noticeably, \emph{LED Off Missing}, \emph{Light Off Missing}, \emph{Matrix Off
Missing}, an \emph{Motor Off Missing}, which are amongst the most frequent bug
patterns (cf. RQ2), always lead to failures. The only fix to stop a robot in
upload mode is to forcefully turn off the robot completely.

Several other bug patterns also always lead to failures (\emph{Colour Out of
Range}, \emph{Low Motor Power}, \emph{Motor Out of Range}, \emph{Several
Launches}, \emph{Useless Light Sensing} and \emph{Useless Line Sensing}). The
\emph{Stuttering Action} and \emph{Useless Distance Sensing} bug patterns each
have one case where the relevant code cannot be reached. More generally, these
bug patterns demonstrate that not knowing the valid value ranges of sensors and
actuators will very likely result in a defect.

The \emph{Sensor Equals Check} bug pattern has the most cases of bugs that had
no observable effects. As the language does not provide a $<=$ operator, users
try to work around this using a disjunction of two comparisons $<$ and $=$.
While the equality is unlikely to be true, this will usually show no effects
thanks to the less-than comparison. \emph{Interrupted Loop Sensing}
produced defects where the interruptions are so short that they are
difficult to observe.

As anticipated, we also found some cases of false positives, where the
programmer used the mechanism we consider as a bug on purpose. \emph{Missing
Loop Sensing} has the highest false positive count (4), here the sensor value at
program start is intentionally used only once to branch off into different
behaviour. In principle, false positives could be avoided by refining the
implementations of the bug patterns to accommodate for these exceptions.

\summary{RQ3}{When bug patterns are executed, they frequently result in
failures, but dead code and safety measures may prevent observable failures.}

\section{Conclusions}\label{sec:conclusions}

Patterns provide a common vocabulary for communicating about code. 
In this paper we demonstrated that \mblock projects are structurally different
from \Scratch programs, and need their own patterns that can deal with the
challenges and possibilities sensors and actuators of robots provide.
To this end we introduced and empirically evaluated a new catalogue of \numbugpatterns bug patterns, three code smells and \numperfumes code perfumes in \mblock. 
Our evaluation found occurrences for all code smells and almost all bug patterns and perfumes, which shows that the concept of bug patterns can be successfully transferred to \mblock.
Furthermore, while some patterns like \emph{Several Launches} are bound to the \mblock environment, patterns based on sensor values and motor power may also be relevant for other block based robot programming languages.

An important next step will be to study the effects of these positive and negative code patterns and the corresponding hints on the learning success of novice programmers, as well as guidelines for instructors on how to teach students about these patterns.
In the future it would be interesting to see if the bug patterns result from misconceptions in programming as some seem to be a symptom of those \cite{swidan2018, sorva2018}.
Our extended version of \litterbox is available at:
\begin{center}
  \url{https://github.com/se2p/LitterBox}
\end{center}

\begin{acks}
	This work is supported by the Federal Ministry of Education and Research
	through the projects 01JA2021 (primary::programming) and 01JA1924 (SKILL.de) as
	part of the ``Qualitätsoffensive Lehrerbildung'', a joint initiative of the
	Federal Government and the Länder. The authors are responsible for the content
	of this publication.
\end{acks}

\balance
\bibliographystyle{ACM-Reference-Format}
\bibliography{references}

\end{document}